\documentclass{emulateapj}

\usepackage{epstopdf}
\usepackage{graphics,graphicx}

\slugcomment{To appear in ApJ}

\begin{document}

\title{Epicyclic Motions and Standing Shocks in Numerically Simulated Tilted Black-Hole Accretion Disks}

\author{P. Chris Fragile}
\affil{Department of Physics \& Astronomy, College of Charleston,
Charleston, SC 29424}
\email{fragilep@cofc.edu}

\and

\author{Omer M. Blaes}
\affil{Department of Physics, University of California, Santa Barbara, CA 93106}

\date{{\small    \today}}
\date{{\small   \LaTeX-ed \today}}

\begin{abstract}
This work presents a detailed analysis of the overall flow structure and unique features of the inner region of the tilted disk simulations described in Fragile et al. (2007). The primary new feature identified in the main disk body is a latitude-dependent radial epicyclic motion driven by pressure gradients attributable to the gravitomagnetic warping of the disk. The induced motion of the gas is coherent over the scale of the entire disk and is fast enough that it could be observable in features such as relativistic iron lines. The eccentricity of the associated fluid element trajectories increases with decreasing radius, leading to a crowding of orbit trajectories near their apocenters. This results in a local density enhancement akin to a compression. These compressions are sufficiently strong to produce a pair of weak shocks in the vicinity of the black hole. These shocks are roughly aligned with the line-of-nodes between the black-hole symmetry plane and disk midplane, with one shock above the line-of-nodes on one side of the black hole and the other below on the opposite side. These shocks enhance angular momentum transport and energy dissipation near the hole, forcing some material to plunge toward the black hole from well outside the innermost stable circular orbit. A new, extended simulation, which was evolved for more than a full disk precession period, allows us to confirm that these shocks and the previously identified ``plunging streams'' precess with the disk in such a way as to remain aligned relative to the line-of-nodes, as expected based on our physical understanding of these phenomena. Such a precessing structure would likely present a strong quasi-periodic signal.

\end{abstract}


\section{Introduction}
\label{sec:intro}

The purpose of this paper is to present a more detailed analysis of
the numerical simulation results described in \citet{fra07b}
(hereafter Paper I). That paper described a global numerical
simulation of an accretion disk subject to the magneto-rotational
instability (MRI) that was misaligned (tilted) with respect to the
rotation axis of a modestly fast rotating ($a/M=0.9$) Kerr black
hole. It was the first such numerical simulation of a tilted disk to
fully incorporate the effects of the black hole spacetime and not
rely on an {\em ad hoc} prescription of angular momentum transport.

A clear identification of a tilted black-hole accretion disk in Nature has yet to be made. Nevertheless, there is reason to believe they may be quite common. In any accretion disk system, the orientation of the disk is set by the net angular momentum of the gas reservoir on large scales. The orientation of the black hole, on the other hand, can either be set by its formation or its evolution. For stellar-mass black holes in low-mass binary systems, only the formation is likely to matter (any evolution subsequent to the formation is unlikely to change the orientation of the black hole significantly). Since the formation of the black hole is largely independent of the angular momentum of the gas reservoir, alignment should generally not be expected in this case. For high mass binaries and active galactic nuclei (AGN), on the other
hand, a large episode of misaligned accretion could reorient the spin of
the black hole \citep{nat99}.  Therefore a more detailed
understanding of the evolutionary history of the system, including 
mergers
in the case of AGN, would be needed to know if a tilted configuration is
expected.

A tilted disk is subject to differential warping due to the effect of Lense-Thirring precession. In disks with a large ``viscous'' stress (parametrized by the \citet{sha73} $\alpha$ parameter) the warping results in the Bardeen-Petterson configuration \citep{bar75,kum85}, characterized by an
alignment of the disk with the equatorial plane of the black hole inside some characteristic warp radius. The Bardeen-Petterson effect has been invoked by a number of authors to explain peculiar observations, such as misaligned jets in AGN \citep{kon05,cap06,cap07} and X-ray binaries \citep{fra01a,mac02}.

However, we did not see evidence for the Bardeen-Petterson effect in our simulation in Paper I. This
was not surprising since our simulation was carried out in the
low stress regime,
with $\alpha < H/r$, where $H$ is the half-height of the disk. 
In this limit warps produced in the disk propagate as waves
\citep{pap95a}, rather than diffusively as in the Bardeen- Petterson
case. Instead of a smooth transition between an untilted disk at small radii and tilted disk at large radii, as with the Bardeen-Petterson effect, we found the tilt to be a long-wavelength oscillatory function of radius. 

The tilted simulation in Paper I also showed other dramatic
differences from comparable simulations of untilted disks. Accretion
onto the hole occurred predominantly through two opposing plunging
streams that started from high latitudes with respect to both the
black-hole and disk midplanes.
These plunging streams also started from a larger radius than the
innermost stable circular orbit (ISCO), which is often assumed to
represent the inner edge of untilted disks.  We interpreted this as
being due to the fact that the tilted disk encounters a
generalized ISCO surface at a larger cylindrical radius than an
untilted disk \citep{fra07a}. In this regard the tilted black hole
effectively acts like an untilted black hole of lesser spin.

Because of the fast sound-crossing time in the disk, the torque of
the black hole acted globally rather than differentially. Instead of
strongly warping the disk, the torque caused the disk to experience
global (solid-body) precession. The precession had a frequency of $\nu_\mathrm{prec} = 3
(M_\odot/M)$ Hz, a value consistent with many observed low-frequency
quasi-periodic oscillations (QPOs). However, this value is strongly
dependent on the size of the disk ($\nu_\mathrm{prec} \propto r_o^{-5/2}$, where $r_o$ is the outer radius), so this frequency may be expected
to be vary over a large range. In the limit of a very large disk or in a case where the disk is strongly coupled to a gas reservoir at large radii, this precession frequency may be
expected to drop to zero.

In the present work we expound on some of the features of the
tilted-disk simulation which were not described in Paper I. This
paper is organized as follows: In \S \ref{sec:methods} we briefly
review the details of the simulations presented in Paper I, in
particular focusing on models 90h (untilted) and 915h (tilted),
which are identical other than the initial tilt of the black hole
relative to the disk ($\beta_0=0$ and $15^\circ$, respectively), and
model 915m, which is a lower resolution version of 915h. In \S
\ref{sec:bending}, we describe the large-scale epicyclic motion that
appears in the tilted simulation but is absent in the comparable
untilted simulation. This epicyclic motion is driven by radial pressure gradients associated with the stationary bending
wave created by the warping action of the gravitomagnetic torque of
the black hole (e.g. \citealt{nel99}).
Next, in \S \ref{sec:shock}, we describe a pair of standing shocks
that again have no counterparts in
untilted simulations. The shocks form roughly
along the line-of-nodes between the disk midplane and black-hole
symmetry plane, one shock on each side of the
black hole. We verify that this is not a chance alignment by following the long-term evolution of
model 915m and confirming that the shocks remain aligned with the line-of-nodes as it precesses with the disk. Finally, in \S \ref{sec:discussion}, we consider some of
the implications of our discoveries, particularly in the context of
relativistic iron lines.

\section{Numerical Methods}
\label{sec:methods}

The simulations in Paper I were carried out using the Cosmos++
astrophysical magnetohydrodynamics (MHD) code \citep{ann05}.
Cosmos++ includes several schemes for solving the GRMHD equations;
in Paper I, the artificial viscosity formulation was used. The
magnetic fields were evolved in an advection-split form, while using
a hyperbolic divergence cleanser to maintain an approximately
divergence-free magnetic field. The GRMHD equations were evolved in
a ``tilted'' Kerr-Schild polar coordinate system
$({t},{r},{\vartheta},{\varphi})$. This coordinate system is related
to the usual (untilted) Kerr-Schild coordinates
$({t},{r},{\theta},{\phi})$ through a simple rotation about the
${y}$-axis by an angle $\beta_0$, as described in \citet[][see also
Fragile \& Anninos 2007]{fra05b}.

The simulations were carried out on a spherical polar mesh with
nested resolution layers. The base grid contained $32^3$ mesh zones
and covered the full $4\pi$ steradians. Varying levels of refinement
were added on top of the base layer; each refinement level doubling
the resolution relative to the previous layer. The main simulations,
models 90h and 915h from Paper I, had two levels of refinement, thus
achieving peak resolutions equivalent to a $128^3$ simulation. To
demonstrate convergence, in Paper I we also presented results at
higher and lower resolutions. In this paper we also discuss results
of model 915m, which used a single level of refinement and had a
peak resolution equivalent to a $64^3$ simulation, but was run for
significantly longer.

In the radial direction a logarithmic coordinate of the form $\eta
\equiv 1.0 + \ln (r/r_{\rm BH})$ was used, where $r_{\rm BH}=1.43 r_G$ is the black-hole horizon radius and $r_G=GM/c^2$ is the gravitational radius. The spatial resolution
near the black-hole horizon was $\Delta r \approx 0.05 r_G$; near
the initial pressure maximum of the torus, the resolution was
$\Delta r \approx 0.5 r_G$. In the angular direction, in addition to
the nested grids, a concentrated latitude coordinate $x_2$ of the
form $\vartheta = x_2 + \frac{1}{2} (1 - h) \sin (2 x_2)$ was used
with $h = 0.5$, which concentrates resolution toward the midplane of
the disk. As a result $r_{\rm center} \Delta \vartheta = 0.3 r_G$
near the midplane while it is a factor of $\sim 3$ larger for the
fully refined zones near the pole. This grid is shown in Figure 1 of
Paper I.

The simulations started from the analytic solution for an
axisymmetric torus around a rotating black hole \citep{cha85}. The
initial torus was identical to model KDP of \citet{dev03c}, which is
the relativistic analog of model GT4 of \citet{haw00}. As in model
KDP, the spin of the black hole was $a/M=0.9$; the inner radius of
the torus was $r_{\rm in}=15 r_G$; the radius of the initial
pressure maximum of the torus $r_{\rm center}=25 r_G$; and the
power-law exponent used in defining the initial specific angular
momentum distribution was $q=1.68$. An adiabatic equation of state
was assumed, with $\Gamma=5/3$. The torus was seeded with a weak
dipole magnetic field in the form of poloidal loops along the
isobaric contours within the torus. The field was normalized such
that initially $\beta_{\rm mag} =P/P_B \ge \beta_{\rm mag,0}=10$
throughout the torus. For the tilted simulations (915h and 915m) the black
hole was inclined by an angle $\beta_0=15^\circ$ relative to the
disk (and the grid) through a transformation of the Kerr metric.
From this starting point, simulations 90h and 915h were allowed to
evolve for a time equivalent to 10 orbits at the initial pressure
maximum, $r_{\rm center}$, corresponding to hundreds of orbits at
the ISCO. Model 915m was evolved for 100 orbital times at the
initial pressure maximum.  Table 1 summarizes the parameters of
all three simulations.

\begin{deluxetable}{cccccc}
\tabletypesize{\scriptsize}
\tablecaption{Simulation Parameters \label{tbl-1}}
\tablewidth{0pt}
\tablehead{
\colhead{Simulation} & \colhead{Tilt} &
\colhead{$a/M$} & \colhead{Equivalent} &
\colhead{$r_{\rm center}$\tablenotemark{a}}
& \colhead{Duration\tablenotemark{b}} \\
\colhead{} & \colhead{Angle} & \colhead{} &
\colhead{Peak} & \colhead{} & \colhead{} \\
\colhead{} & \colhead{} & \colhead{} &
\colhead{Resolution} & \colhead{} & \colhead{} \\
}
\startdata
90h & 0 & 0.9 & $128^3$ & 25 $r_{\rm G}$ & 10 \\
915m & $15^\circ$ & 0.9 & $64^3$ & 25 $r_{\rm G}$ & 100 \\
915h & $15^\circ$ & 0.9 & $128^3$ & 25 $r_{\rm G}$ & 10 \\
\enddata

\tablenotetext{a}{Radius of initial pressure maximum.}
\tablenotetext{b}{In units of $t_{\rm orb}$, the geodesic orbital period
at the initial pressure maximum $r_{\rm center}$.}

\end{deluxetable}

\section{Epicyclic Motion}
\label{sec:bending}

As noted above, Lense-Thirring precession causes differential warping in tilted black-hole accretion disks. In the thick-disk regime, appropriate for the simulations described in Paper I, warp disturbances are expected to propagate in a wave-like, rather than diffusive, manner. For nearly Keplerian disks, such as those resulting from MRI turbulent simulations, the resulting bending wave is expected to produce large horizontal motion within the disk \citep{nel99,tor00}. The radial and azimuthal components of this motion are odd functions of $z$, leading to large vertical shear ($\partial V^r/\partial z$, $\partial V^\phi/\partial z$). 

\begin{figure*}
\includegraphics[scale=.6]{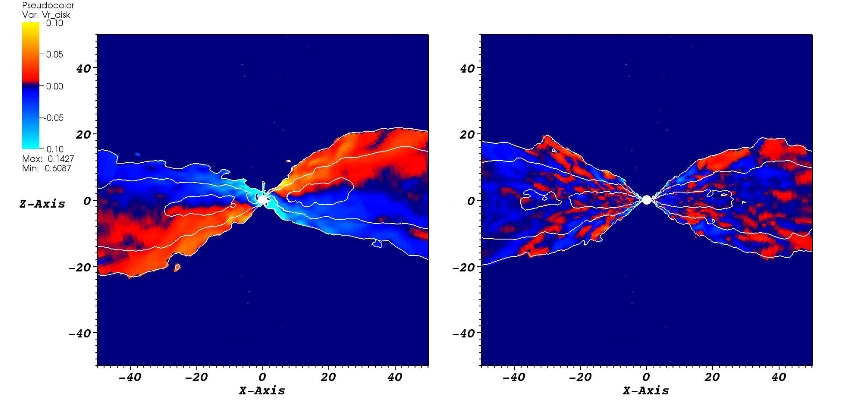} \caption{Meridional plots
($\varphi=0$) through the final dumps ($t=10t_\mathrm{orb}$) of
simulations 915h ({\em left}) and 90h ({\em right}) showing a
pseudocolor representation of $V^r$ for outflowing ($V^r > 0$) and
inflowing ($V^r < 0$) gas as {\em hot} and {\em cold} colors,
respectively. The velocity scale is normalized to the speed of light
$c$. The plots are overlaid with isocontours of density at $\rho =
0.4$, 0.04, and $0.004\rho_\mathrm{max,0}$.
Material with $\rho<0.004\rho_\mathrm{max,0}$ is excluded from the
figure. The figure is oriented in the sense of the grid, such that
the black hole is tilted $15^\circ$ to the left in the {\em left}
panel. By this time in simulation 915h the disk has precessed such
that the angular momentum axis of the disk is no longer in the plane
of this figure. At this point $\mathbf{J}_\mathrm{disk}$ has
acquired a positive $y$-component and a negative $x$-component.
\label{fig:rhoVr1}}
\end{figure*}
\begin{figure*}
\includegraphics[scale=.66]{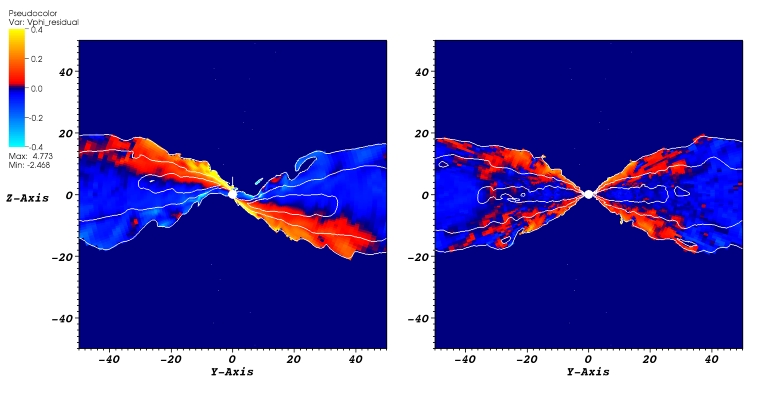}\caption{Meridional plots
($\varphi=90^\circ$) through the final dumps ($t=10t_\mathrm{orb}$)
of simulations 915h ({\em left}) and 90h ({\em right}) showing a
pseudocolor representation of $(V^\phi-\Omega)/\Omega$ for
superorbital ($V^\phi
> \Omega$) and suborbital ($V^\phi < \Omega$) gas as {\em hot} and
{\em cold} colors, respectively, where $\Omega = (M/r^3)^{1/2}/[1+a(M/r^3)^{1/2}]$ is the particle orbital angular frequency. The plots are overlaid with
isocontours of density at $\rho = 0.4$, 0.04, and
$0.004\rho_\mathrm{max,0}$. Material with
$\rho<0.004\rho_\mathrm{max,0}$ is excluded from the figure. The
figure is oriented in the sense of the grid, such that the black
hole is tilted $15^\circ$ away from the viewer's vertical in the {\em left}
panel. \label{fig:rhoVph1}}
\end{figure*}

Figures \ref{fig:rhoVr1} and \ref{fig:rhoVph1} compare the radial and azimuthal motions of the gas for our tilted and untilted simulation (915h and 90h). Clearly
the fluid velocity in the tilted disk is no longer dominated by
turbulent motion as it is for untilted disks \citep{dev03b}, but has
an ordered sense about it. For instance, in Figure
\ref{fig:rhoVr1}{\em a} we see that in the right-hand side of the
image the gas in the upper layers of the disk is moving radially
outward, whereas the gas in the lower layers is moving radially
inward. The sense of motion is reversed on the left-hand side of the
image. Similarly, on the right-hand side of Figure
\ref{fig:rhoVph1}{\em a} the gas in the upper half of the disk is
moving slower than the bulk angular velocity, whereas the gas in the
lower half is moving faster than this average. Again, the sense of
the flow is reversed on the left-hand side of the image. This
pattern of motion is reminiscent of epicyclic motion, with the upper
and lower halves of the disk executing epicycles that are
$180^\circ$ out of phase with one another. Notice that no such
organized motion is apparent in our untilted simulation (Figures
\ref{fig:rhoVr1}{\em b} and \ref{fig:rhoVph1}{\em b}).

\begin{figure*}
\includegraphics[scale=.66]{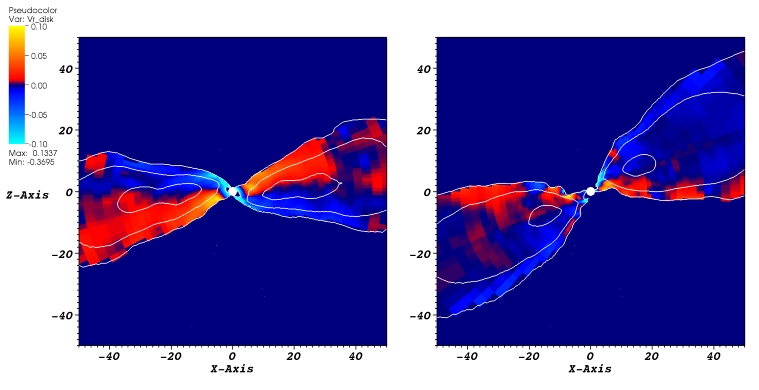} \caption{Meridional plots
($\varphi=0$) at times $t=10$ ({\em left}) and $50t_\mathrm{orb}$ ({\em right})
(representing approximately 1/2 precession period difference) from simulation
915m showing a
pseudocolor representation of $V^r$ for outflowing ($V^r > 0$) and
inflowing ($V^r < 0$) gas as {\em hot} and {\em cold} colors,
respectively. The color scale and contours are normalized the same as in Fig. \ref{fig:rhoVr1}. The figure is oriented in the sense of the grid, such that
the black hole is tilted $15^\circ$ to the left. Note that the sense of the radial motion as seen from
this fixed viewing direction is reversed between the two frames. This is consistent with the epicyclic motion tracking the precession of the disk.
\label{fig:rhoVr2}}
\end{figure*}

\begin{figure*}
\includegraphics[scale=.66]{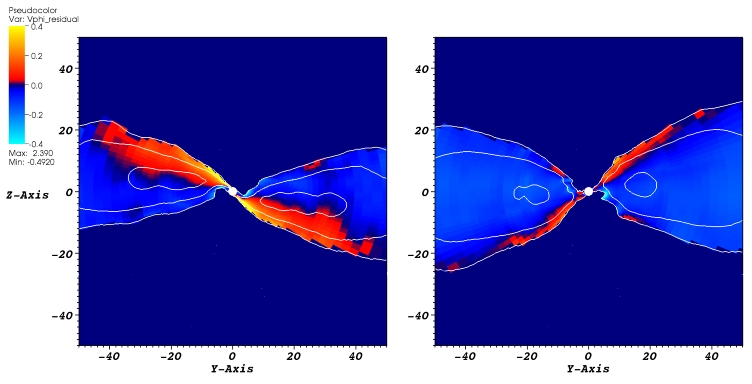} \caption{Meridional plots
($\varphi=90^\circ$) at times $t=10$ ({\em left}) and $50t_\mathrm{orb}$ ({\em right})
(representing approximately 1/2 precession period difference) from simulation
915m showing a
pseudocolor representation of $(V^\phi-\Omega)/\Omega$ for
superorbital ($V^\phi
> \Omega$) and suborbital ($V^\phi < \Omega$) gas as {\em hot} and
{\em cold} colors, respectively. The color scale and contours are normalized the same as in Fig. \ref{fig:rhoVph1}. The
figure is oriented in the sense of the grid, such that the black
hole is tilted $15^\circ$ away from the viewer's vertical.
Note that the sense of the residual azimuthal motion as seen from this fixed viewing direction is
reversed between the two frames. This is consistent with the epicyclic motion tracking the precession of
the disk. \label{fig:rhoVph2}}
\end{figure*}

We find that the pattern of the epicyclic motion is tied to
the orientation of the disk relative to the black hole. As our
simulated tilted disk 915m precesses, the velocity pattern of the epicyclic
motion changes as represented in Figures \ref{fig:rhoVr2} and
\ref{fig:rhoVph2}, where we show the velocity patterns at $t=10$ and
$t=50t_\mathrm{orb}$, approximately 1/2 precession period apart.
Note that the sense of the epicyclic motion is reversed between the
two different evolution times.

\begin{figure}
\begin{center}
\includegraphics[scale=.675]{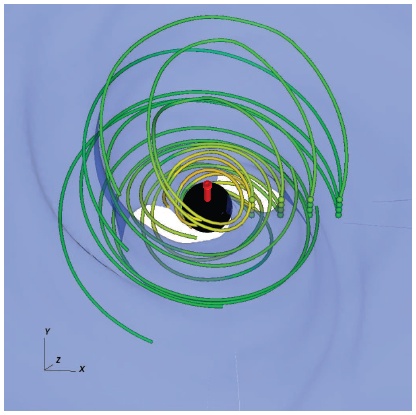} \caption{Isosurface plot
of density ({\em semitransparent blue}) plus selected streamlines for simulation 915h. The
isosurface at $\rho=0.1\rho_\mathrm{max,0}$ is chosen to highlight
the plunging streams. There are 16 streamlines that begin in a $4
\times 4$ lattice in the $xz$-plane. The lattice is centered about
the symmetry plane of the black hole and the figure is oriented
looking almost directly down the spin axis of the hole (red arrow). The spin axis is actually tilted $10^\circ$ away from the viewer's line-of-sight to offer a better perspective.
\label{fig:rho_stream}}
\end{center}
\end{figure}

Figure \ref{fig:rho_stream} shows the epicyclic motion caused by the
bending wave from a different perspective. In the figure, a
$4\times4$ lattice of streamlines starts in the $xz$-plane. The
lattice is centered above and below the symmetry plane of the black
hole and the viewer is looking almost down the black-hole spin axis (the black-hole spin axis is tilted $10^\circ$ away from the viewer's line-of-sight to give a better perspective). As
expected from the previous figures of simulation 915h at
$t=10t_\mathrm{orb}$, the streamlines that begin above the symmetry
plane of the disk are initially moving radially outward, whereas
those that begin below are moving radially inward. After
approximately one-quarter of an orbit, the upper streamlines have
reached their apocenter and the radial motion changes direction.
This is also where material encounters one of the standing shocks,
described in the next section, which explains the sudden change in
direction of the streamlines. Much of this material then begins
plunging toward the hole, ultimately accreting within a couple
orbits. Thus, as noted in \citet{fra05b}, high-latitude material is
preferentially being drained from the disk.

Figure \ref{fig:rho_stream} also reveals that the eccentricity $e$ of the particle streamlines depends on the height of the streamline above the disk midplane (the top streamlines are more eccentric that those one row below). This is consistent with the expectation that $e \propto \xi$ \citep{iva97}, where $\xi$ is the vertical distance measured perpendicular to the midplane of the disk ($\xi$ is the $z$-component of a cylindrical coordinate system ($R$, $\psi$, $\xi$) aligned with each concentric ring of the disk, what is sometimes called the twisting coordinate system).

\section{Standing Shocks}
\label{sec:shock}

\begin{figure}
\plotone{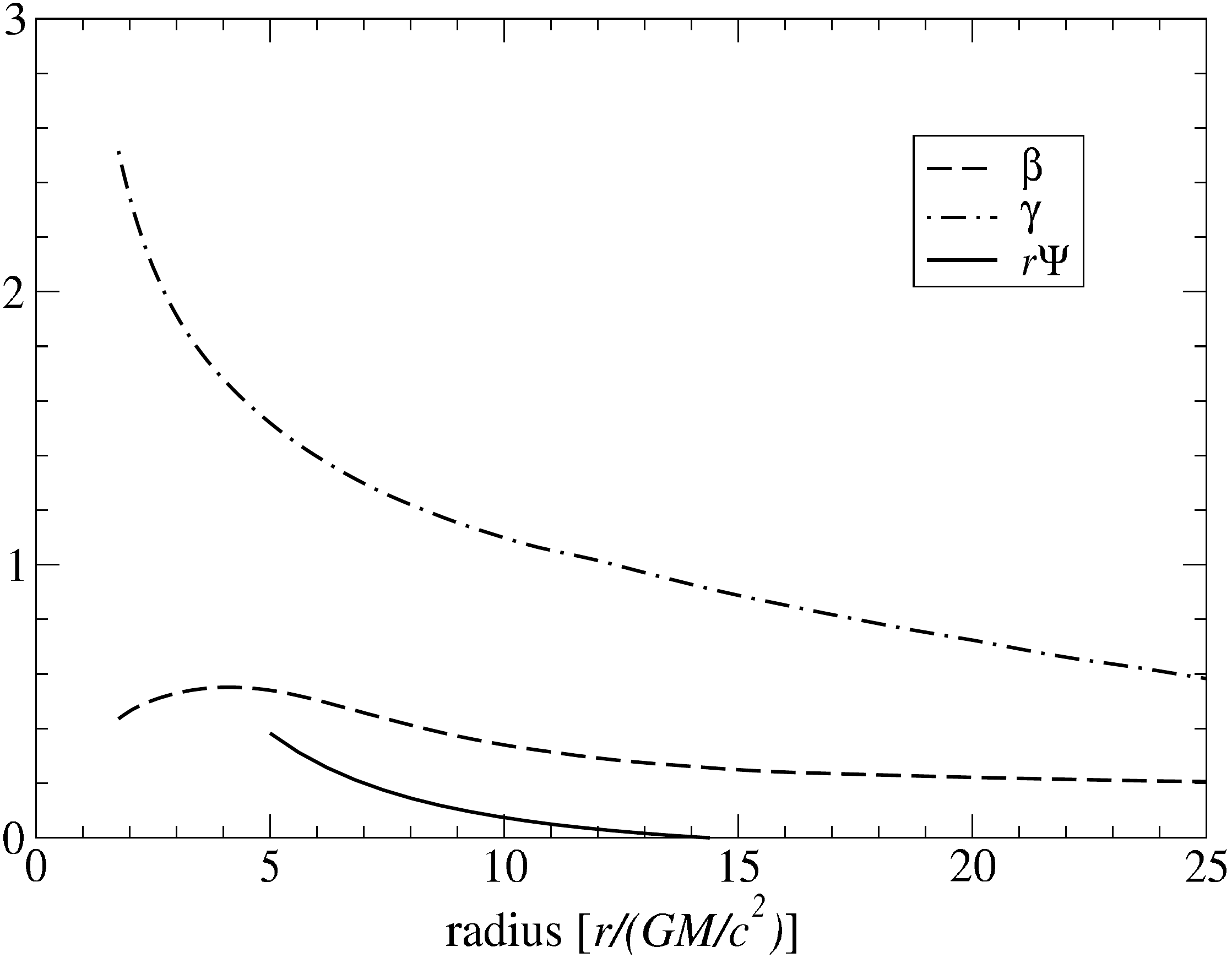} \caption{Plot of the tilt $\langle\beta \rangle_t$, twist $\langle \gamma \rangle_t$, and radial dependence of the orbital eccentricity $r\Psi$ as a function of radius through the disk. The data for this plot has been time-averaged from $t=9t_\mathrm{orb}$ to $10t_\mathrm{orb}$. The initial tilt and twist were 0.2618 and 0, respectively. \label{fig:eccentricity}}
\end{figure}

The full expression for the eccentricity of the orbit of each fluid element is \citep{iva97}
\begin{equation}
e = \frac{R\xi}{6M} \Psi = \frac{R\xi}{6M} \frac{\partial(\beta \cos \gamma)}{\partial R} ~,
\end{equation}
where $\beta$ and $\gamma$ are the tilt and twist of the disk, respectively, defined for each concentric ring. In Figure \ref{fig:eccentricity} we plot $\beta(r)$, $\gamma(r)$, and $r\Psi(r)$ (for computational convenience we have replaced the cylindrical twisting coordinate $R$ with the spherical-polar radius $r$). It is important to note the increase in $r\Psi$ with decreasing $r$. Such behavior, with orbits getting more eccentric closer to the black hole, results in a concentration of fluid element trajectories near their respective apocenters \citep[see Fig. 5a of][]{iva97}. Notice in Figure \ref{fig:rho_stream}, for instance, that each of the top four streamlines makes its closest approach to the next streamline out when it is at its own apocenter.

\begin{figure}
\begin{center}
\includegraphics[scale=.298]{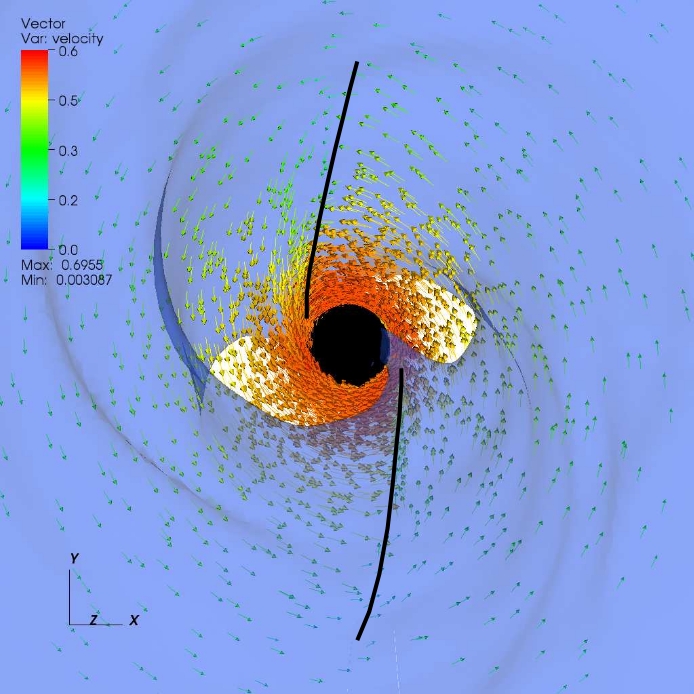} \caption{Isosurface plot of
density ({\em semitransparent blue}) plus selected velocity vectors for simulation 915h at
$t=10t_\mathrm{orb}$. The density isosurface is the same as in Fig.
\ref{fig:rho_stream}. The figure is oriented
looking directly down the spin axis of the hole. Velocity vectors are only included for gas
within $\pm 0.5 r_G$ of the initial disk midplane to prevent
overcrowding the image. The sudden change in direction and magnitude
of the velocity vectors (highlighted by the black lines) is indicative of a shock.
\label{fig:rhoV}}
\end{center}
\end{figure}

The crowding of fluid element trajectories near their apocenters produces a local density enhancement. This will be most pronounced away from the disk midplane because of the dependence of $e$ on $\xi$. Because the fluid elements are traveling supersonically, this local density enhancement produces a weak shock in the flow. In Figure \ref{fig:rhoV} we
identify the standing shocks by the sudden change in the magnitude
and direction of the velocity vectors along a roughly linear feature
associated with the leading edge of the each plunging stream.

\begin{figure}
\begin{center}
\includegraphics[scale=.298]{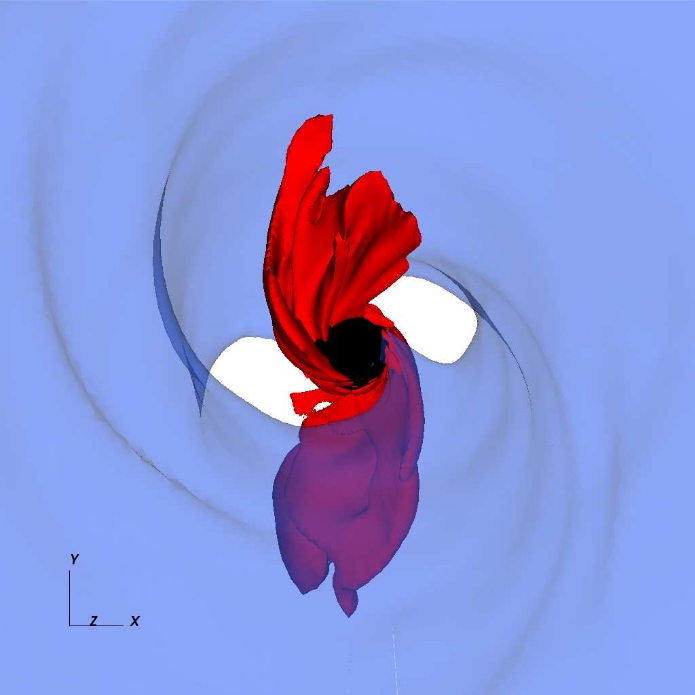} \caption{Isosurface plots
of density ({\em semitransparent blue}) and $\vert\mathrm{curl~}
\mathbf{V}\vert$ ({\em red}) for simulation 915h at
$t=10t_\mathrm{orb}$. $\vert\mathrm{curl~} \mathbf{V}\vert$ is a
good tracer of the location of a shock. The plot is restricted to
the region $\pi/4<\vartheta<3\pi/4$ to prevent overcrowding of the
image from shocks associated with the outflowing jets. The density
isosurface is the same as in Fig. \ref{fig:rho_stream}. The figure is oriented
looking directly down the spin axis of the hole.
\label{fig:rho_curlV}}
\end{center}
\end{figure}

Another way to identify the standing shocks is by plotting the
magnitude of the vorticity, indicated by $\vert\mathrm{curl~}
\mathbf{V}\vert = \vert\nabla \times \mathbf{V}\vert$, which
increases at a shock. Figure \ref{fig:rho_curlV} shows an isosurface
of $\vert\mathrm{curl~} \mathbf{V}\vert$ overlaid on a density
isosurface. Very similar results are obtained if the gradient of
the entropy, another indicator of the presence of a shock, is
plotted instead. Notice that one of the shock surfaces lies mostly above
the chosen isodensity surface, while the other lies below it.
This is consistent with the linear dependence of $e$ on $\xi$. We don't expect the shocks to extend into the disk midplane where epicyclic motion ceases. The location of the shocks in azimuth is also consistent with the apocenters of the fluid elements trajectories, as inferred from Figures \ref{fig:rhoVr1} -- \ref{fig:rho_stream}.

\begin{figure}
\begin{center}
\includegraphics[scale=.298]{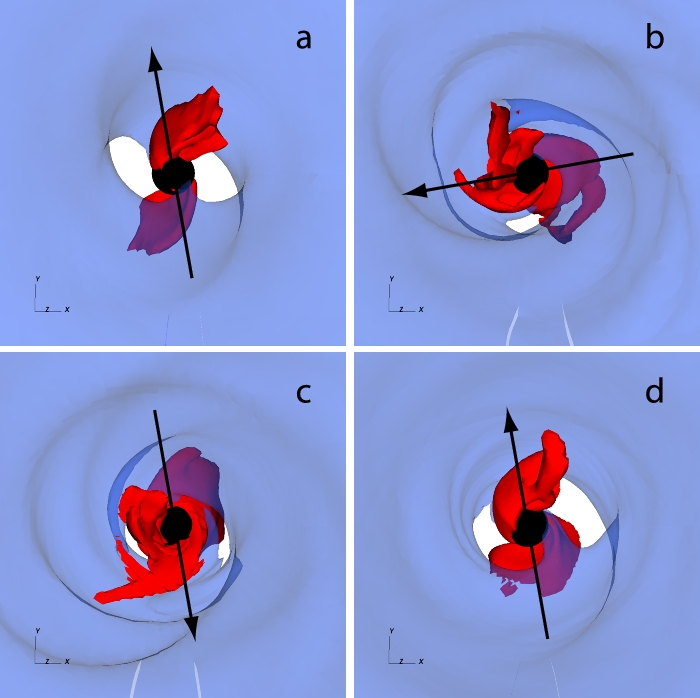} \caption{Same as Fig.
\ref{fig:rho_curlV} except at times $t=10$ (panel {\em a}), 32
(panel {\em b}), 50 (panel {\em c}), and $100t_\mathrm{orb}$ (panel
{\em d}) (representing approximately 0.1, 0.35, 0.6, and 1.1
precession period) from simulation 915m. Also, the isodensity
surfaces are at $\rho=0.04$, 0.01, 0.02, and
$0.02\rho_\mathrm{max,0}$, respectively, instead of
$0.1\rho_\mathrm{max,0}$ as in previous figures. The arrows are
included to give a visual reference of the precession of  the
line-of-nodes. By noting which shock is above the black-hole
symmetry plane and which is below we see that: the shocks in panels
{\em a} and {\em b}, {\em a} and {\em c}, and {\em a} and {\em d}
are $90^\circ$, $180^\circ$, and $360^\circ$ out of phase,
respectively. It is apparent that both the plunging stream and shock
precess with the disk. \label{fig:915m_shock}}
\end{center}
\end{figure}

Based on our physical understanding of the plunging streams identified in Paper I and the standing shocks identified in this paper, we expect both features to maintain a constant orientation vis a vis the line-of-nodes between the disk midplane and black-hole symmetry plane. To confirm this we have evolved our ``medium''
resolution simulation (model 915m of Paper I) for
$100t_\mathrm{orb}$, which is more than a full precession period.
Figure \ref{fig:915m_shock} shows the plunging streams and standing
shocks at 0.1, 0.35, 0.6, and 1.1 precession periods. The
orientations of both features closely track the precession of the
disk and maintain roughly constant alignments relative to the line-of-nodes. 

\subsection{Shock Effects}
\begin{figure}
\plotone{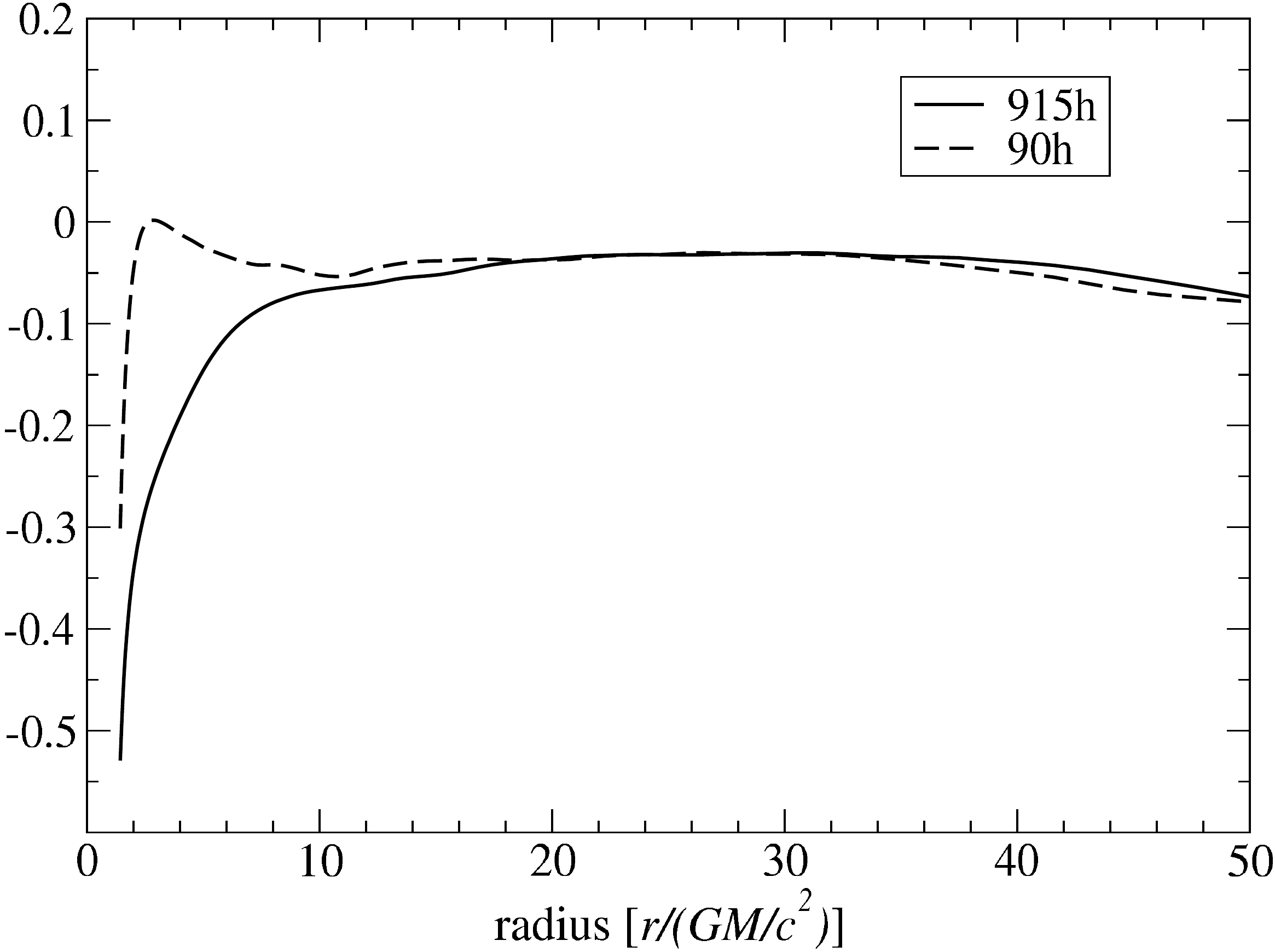} \caption{Plot of the weighted residual
specific angular momentum $(\langle\langle\ell\rangle_A\rangle_t -
\ell_\mathrm{cir})/\ell_\mathrm{cir}$ as a function of radius for
simulations 915h ({\em solid line}) and 90h ({\em dashed line}). The
data has been time-averaged over the interval $t=7$ to $10t_{\rm
orb}$. The residuals are calculated from the specific angular
momenta of circular orbits with inclinations of $15^\circ$ and
$0^\circ$, respectively. \label{fig:ang_mom}}
\end{figure}

The presence of non-axisymmetric standing shocks in the accretion
disk can have important consequences, principally including enhanced
angular momentum transport and dissipation. To
illustrate the former, in Figure \ref{fig:ang_mom} we plot angular
momentum residual weighted by the circular orbit value, i.e. we plot
$(\langle\langle\ell\rangle_A\rangle_t -
\ell_\mathrm{cir})/\ell_\mathrm{cir}$ as a function of radius, where
$\ell_\mathrm{cir}$ is the angular momentum of circular orbits with
inclinations of $15^\circ$ and $0^\circ$ for simulations 915h and 90h, respectively, calculated
from the following expression (equation 26 of Paper I)
\begin{equation}
\ell = \frac{N_1 + \Delta (Mr)^{1/2} N_2^{1/2} \cos i}{D} ~,
\end{equation}
where
\begin{equation}
N_1 = -aMr \left(3r^2 + a^2 - 4Mr \right) \cos^2 i ~,
\end{equation}
\begin{equation}
N_2 = r^4 + a^2 \sin^2 i \left(a^2 + 2r^2 - 4Mr \right) ~,
\end{equation}
\begin{equation}
D = a^2 \left( 2r^2 + a^2 - 3Mr \right) \sin^2 i + r^4 + 4M^2r^2 -
4r^3M - Mra^2 ~,
\end{equation}
and
\begin{equation}
\Delta=r^2-2Mr+a^2 ~;
\end{equation}
and $\langle\ell\rangle_A =
\langle\rho\ell\rangle_A/\langle\rho\rangle_A$ is the
density-weighted shell average of the specific angular momentum. 
Shell averaged quantities are computed as:
\begin{equation}
\langle\mathcal{Q}\rangle_A(r,t) = \frac{1}{A} \int^{2\pi}_0
\int^{\vartheta_2}_{\vartheta_1} \mathcal{Q} \sqrt{-g}
\mathrm{d}\vartheta \mathrm{d}\varphi ~,
\end{equation}
where $A = \int^{2\pi}_0 \int^{\vartheta_2}_{\vartheta_1} \sqrt{-g}
\mathrm{d}\vartheta \mathrm{d}\varphi$ is the surface area of the
shell. The data have also been time-averaged over the interval,
$7t_{\rm orb} = t_{\rm min} \le t \le t_{\rm max} = 10t_{\rm orb}$,
where time averages are defined as
\begin{equation}
\langle\mathcal{Q}\rangle_t = \frac{1}{t_{\rm max} - t_{\rm min}}
\int^{t_{\rm max}}_{t_{\rm min}} \mathcal{Q} \mathrm{d}t ~.
\end{equation}
The sharp down-turn of the specific angular momentum inside
$r\lesssim 10 r_G$ in simulation 915h is indicative of extra angular
momentum being removed from the flow. This suggests that the
standing shock plays a significant role in the transport of angular
momentum in the tilted disk.

\begin{figure}
\plotone{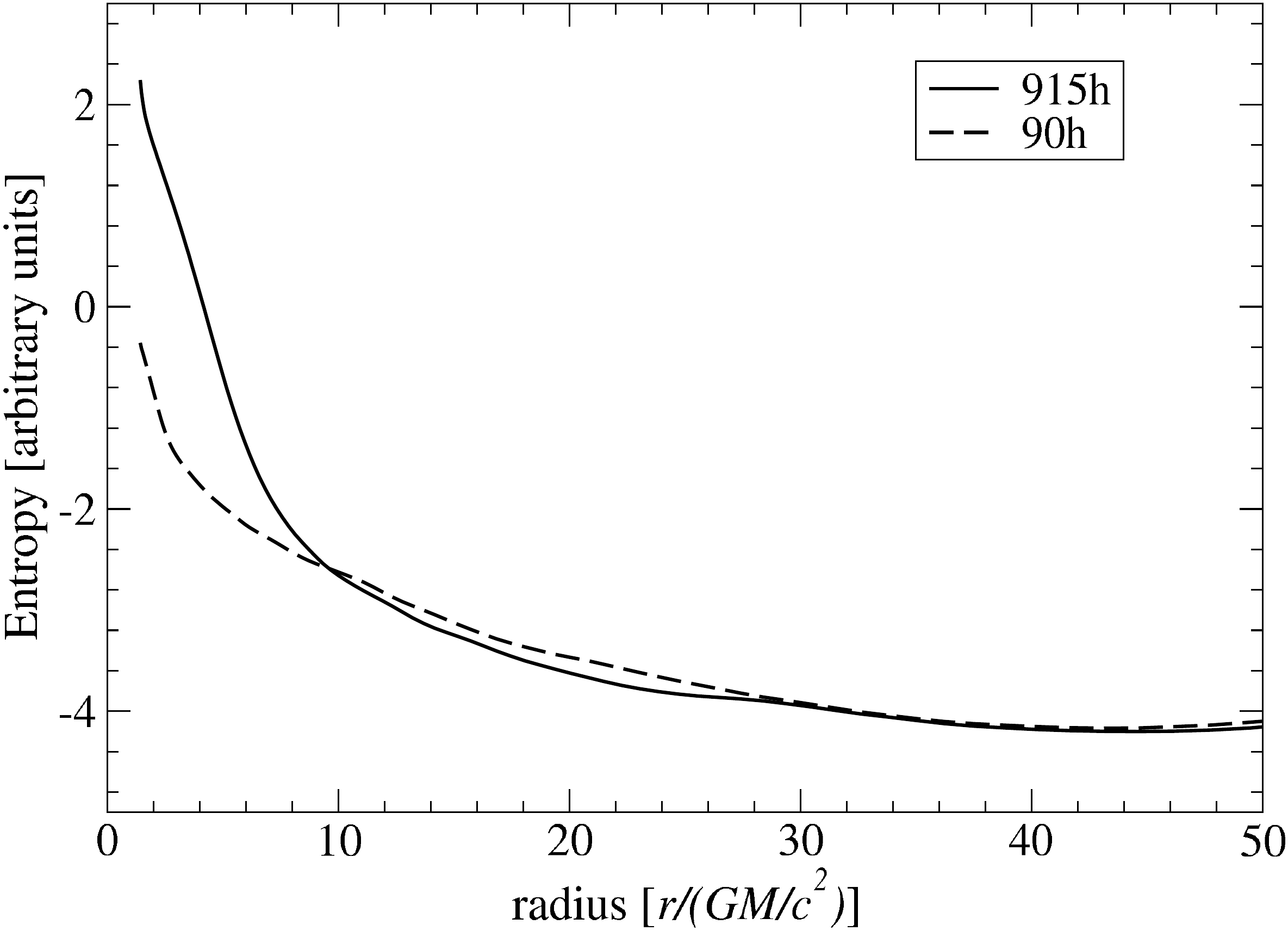} \caption{Plot of the density-weighted
shell average of the fluid entropy as a function of radius for
simulations 915h ({\em solid line}) and 90h ({\em dashed line}). The
data has been time-averaged over the interval $t=7$ to $10t_{\rm
orb}$. \label{fig:entropy}}
\end{figure}

The shock also enhances the energy dissipation in the disk. This is
shown in Figure \ref{fig:entropy}, where we plot the
density-weighted shell averages of the fluid entropy in the tilted
and untilted disks. Actually, since we are only interested in
relative changes in entropy, for simplicity we plot $\mathcal{S} =
\ln (P/\rho^\Gamma)$. The
significant enhancement in entropy generation inside $r \lesssim 10
r_G$ for the tilted disk indicates the extra dissipation provided by the shock. This up-turn nicely coincides with the down-turn in Figure \ref{fig:ang_mom}, further supporting the association of both effects with the shock.

\begin{figure*}[htbp]
\plottwo{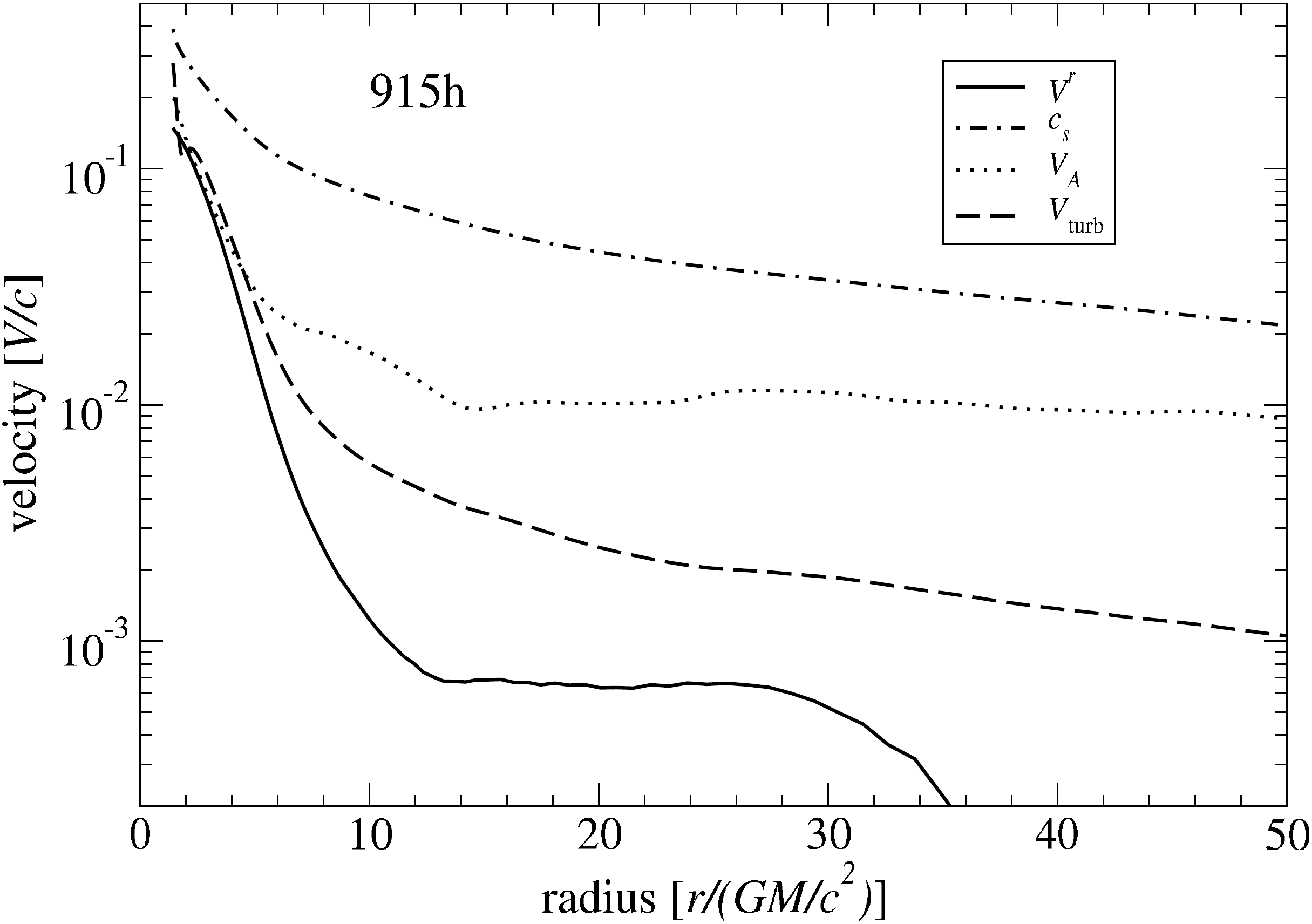}{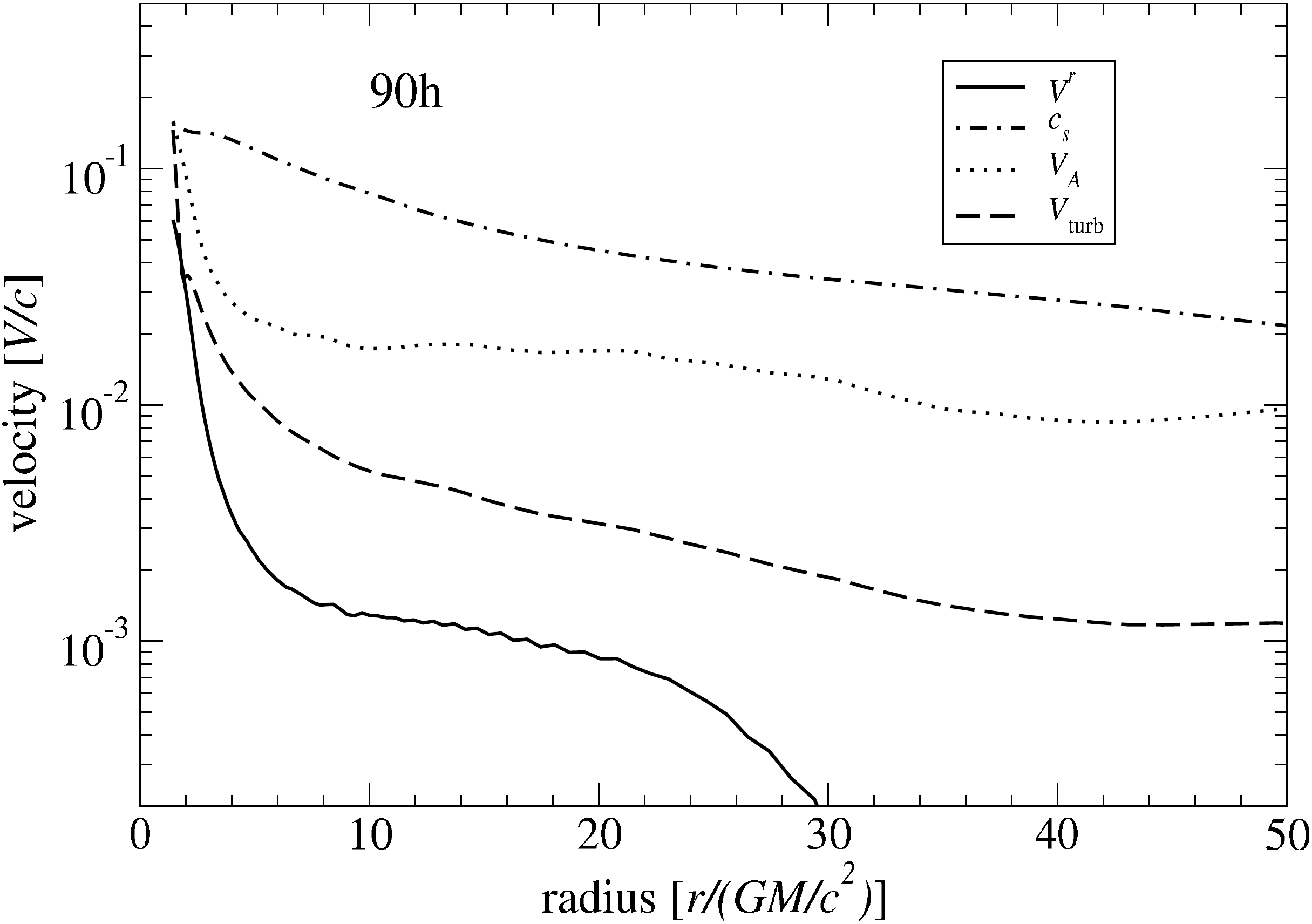} \caption{Plot of
characteristic velocities as functions of radius within the tilted
({\em left panel}) and untilted ({\em right panel}) disks. All data
have been time-averaged over the interval $t=7$ to $10t_{\rm orb}$.
Of particular importance is the sharp upturn in $V^r$, indicating
the start of the plunging region, which happens at a larger radius
for the tilted simulation. \label{fig:velocities}}
\end{figure*}

The extra energy dissipation and angular momentum transport
significantly affect the inner
regions of the disk. In Figure \ref{fig:velocities}, we see that the
radial plunging region, defined by the sharp upturn in
$\overline{V}^r$, begins at a considerably larger radius for the
tilted simulation (915h) than for the untilted (90h). This can have
important implications for disk observations \citep{kro02} since most of the radiated energy from a disk comes from just outside the plunging radius. Of special concern is
the common use of the inner edge of the disk as a direct indicator
of the spin of the black hole. Clearly tilt must be taken into
account if one wants to relate the plunging radius of the disk to
the spin of the black hole for tilted disks. Figure
\ref{fig:velocities} also compares other characteristic velocities
associated with the disks. All of the velocities are density
weighted shell averages $\overline{V}=\langle \langle \rho V
\rangle_A/\langle \rho \rangle_A \rangle_t$. The local sound speed
is recovered from $c_s^2=\Gamma (\Gamma-1)P/[(\Gamma-1)\rho + \Gamma
P]$. The Alfv\'en speed is
\begin{equation}
v_A =  \sqrt{\frac{\vert\vert B \vert\vert^2}{4 \pi \rho h +
\vert\vert B \vert\vert^2}} ~.
\end{equation}
We approximate the turbulent velocity as $V_\mathrm{turb}\approx
\alpha^{1/2}c_s$, where
\begin{equation}
\alpha = \left\langle \frac{ \vert u^r u^\varphi \vert\vert B
\vert\vert^2 - B^r B^\varphi \vert}{4 \pi P} \right\rangle_A
\end{equation}
is the dimensionless stress parameter. The most notable difference
is in the radial inflow velocity $\overline{V}^r$, although there is also some indication that the turbulent stress rises faster in the interior of the tilted disk.


\subsection{Post-Shock Magnetic Field}
\begin{figure}[b]
\begin{center}
\includegraphics[scale=.298]{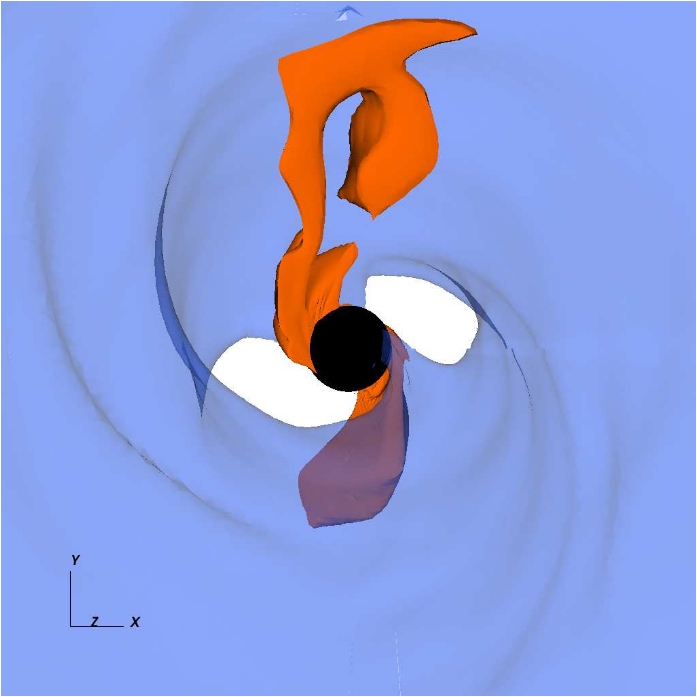} \caption{Isosurface plot
of density ({\em semitransparent blue}) plus an isosurface of $\beta_\mathrm{mag}=P/P_B =
0.1$ ({\em orange}), indicating a thin layer of magnetically dominated plasma behind
the shock. The density isosurface is the same as in Fig. The figure is oriented
looking directly down the spin axis of the hole.
\ref{fig:rho_stream}. \label{fig:beta}}
\end{center}
\end{figure}

We found previously that magnetic fields remain subthermal
everywhere in the disk throughout the simulation (see Figure 11 of
Paper I). They, therefore, must not play a significant role in the
dynamics of the epicyclic motion, plunging streams, or standing shocks.
This statement is further supported by the fact that very similar
features were seen in earlier {\em hydrodynamic} simulations of
tilted black-hole accretion disks \citep{fra05b}. Despite the minor
role of the magnetic fields in the dynamics, the presence of a
standing shock can enhance the strength of the magnetic field. From
\citet{fra05a} we note that the post-shock value of
$\beta_\mathrm{mag}$ in the Newtonian limit can depend sensitively on the strength of the
shock
\begin{equation}
\beta_\mathrm{mag,ps} = 2\Gamma(\Gamma-1)^2/(\Gamma+1)^3
\mathcal{M}^2 \beta_{\mathrm{mag},i}
\end{equation}
where $\beta_{\mathrm{mag},i}$ is the pre-shock value and we assume the field is oriented perpendicular to the shock normal. For
$\Gamma=5/3$, as in our work, this gives
\begin{equation}
\beta_\mathrm{mag,ps} = 0.078 \mathcal{M}^2 \beta_{\mathrm{mag},i}
~.
\end{equation}
Therefore, for $\mathcal{M} < 3.6$, the post-shock
$\beta_\mathrm{mag}$ will be lower than the pre-shock value. In
Figure \ref{fig:beta} we show that there is, indeed, a thin layer of
magnetically dominated plasma just behind the shock. To get
$\beta_\mathrm{mag,ps}=0.1$ as Figure \ref{fig:beta} starting from
$\beta_{\mathrm{mag},i}=1$ (a reasonable guess for the high latitude
material), we need $\mathcal{M}=1.13$, which is consistent with what
we see in the simulation. Again, these do not appear to be
tremendously strong shocks.

\section{Discussion}
\label{sec:discussion}

In this paper we have compared two simulations that are identical to
one another in every respect except one: the initial tilt between
the black-hole and disk angular momenta. Despite the apparent
similarity, we observed remarkable differences in the evolution of
these two simulations.

The primary new feature that we describe in the main disk body is a
strong epicyclic driving attributable to the gravitomagnetic torque
of the misaligned (tilted) black hole. The induced motion of the gas
is coherent over the scale of the entire disk. An interesting point
about this epicyclic motion that has not been made before is that it
could be detectable, for instance in the profile of relativistically
broadened iron lines. We have previously pointed out the importance
of the iron lines in directly probing tilted accretion disks
\citep{fra05c}, but at the time we were not in a position to
recognize the importance of the epicyclic motion. From Figure
\ref{fig:rhoVph1} we get that the velocities associated with the
epicyclic motion represent a significant fraction ($\lesssim40\%$)
of the orbital velocity of the gas. The corresponding shift in a
reflection feature such as the iron line should be of a similar
magnitude. The interesting thing is to note that an observer viewing
the disk from the vantage point of Figure \ref{fig:rhoVph1}{\em a}
and seeing the ``top'' of the disk would see a smaller than expected
red shift (the gas going away is not moving as fast as expected) and
a larger than expected blue shift (the approaching gas is moving
faster than expected). An observer viewing the same disk from the
same vantage point but seeing the ``bottom'' of the disk would see
exactly the opposite shift. Therefore, depending on the viewing
angle, the entire line profile of a misaligned disk could be shifted
toward the blue {\em or} the red relative to an aligned disk. On the
redshifted side this effect might be confused with gravitational
redshifting, making it appear that the line is coming from deeper in
the potential well than is actually the case.

If the disk actually precesses then there is a clear and simple way
to disentangle this effect because the epicyclic motion is phased
with the orientation of the disk, as discussed in \S
\ref{sec:bending} above. This means that the line shift we are
describing would reverse itself twice per precession period,
appearing for half a precession period as an overall blueshift and
for the other half as an overall redshift.
If the integration time of the detector is longer than the
precession period, the net effect would generally be to smear
or broaden the line. This may provide an alternative explanation for an exceptionally broad line profile such as GX 339-4 \citep{mil04}, without requiring a high black-hole spin or small disk radius. On the other hand, if the integration time is
shorter than the precession period, then one would expect the iron line to vary in phase with changes in the X-ray flux (i.e. in conjunction with a low frequency QPO that would correspond to the precession frequency of the disk). Such behavior has been identified in at least one source, GRS 1915+105 \citep{mil05}.

The second new feature of our tilted disk simulation that we
described is a pair of standing shocks, roughly aligned with the
line-of-nodes between the disk and black hole symmetry planes. 
This asymmetric shock provides additional angular momentum transport and energy dissipation in the tilted disk, relative to the untilted one. This enhanced dissipation may help compensate for the loss of radiative efficiency (relative to an untilted disk) due to the plunging region 
starting further out than the equatorial ISCO radius of the spinning black hole. 
In collisionless accretion flows, such as those thought to be relevant for
Sgr~A* \citep[e.g.][]{qua03}, it is conceivable that such shocks could
also be sites of particle acceleration.  The precession of these shocks 
might then result in periodic variations of nonthermal radiation.

In considering the magnetic fields in the inner part of the tilted
disk, as noted in Paper I, the magnetic fields remain largely
subthermal. They do not play an important role in the physics of the
epicyclic motion, plunging streams, or standing shocks, although we do
find some enhancement of $\beta_\mathrm{mag}$ associated with the
shocks.

\begin{figure}
\plotone{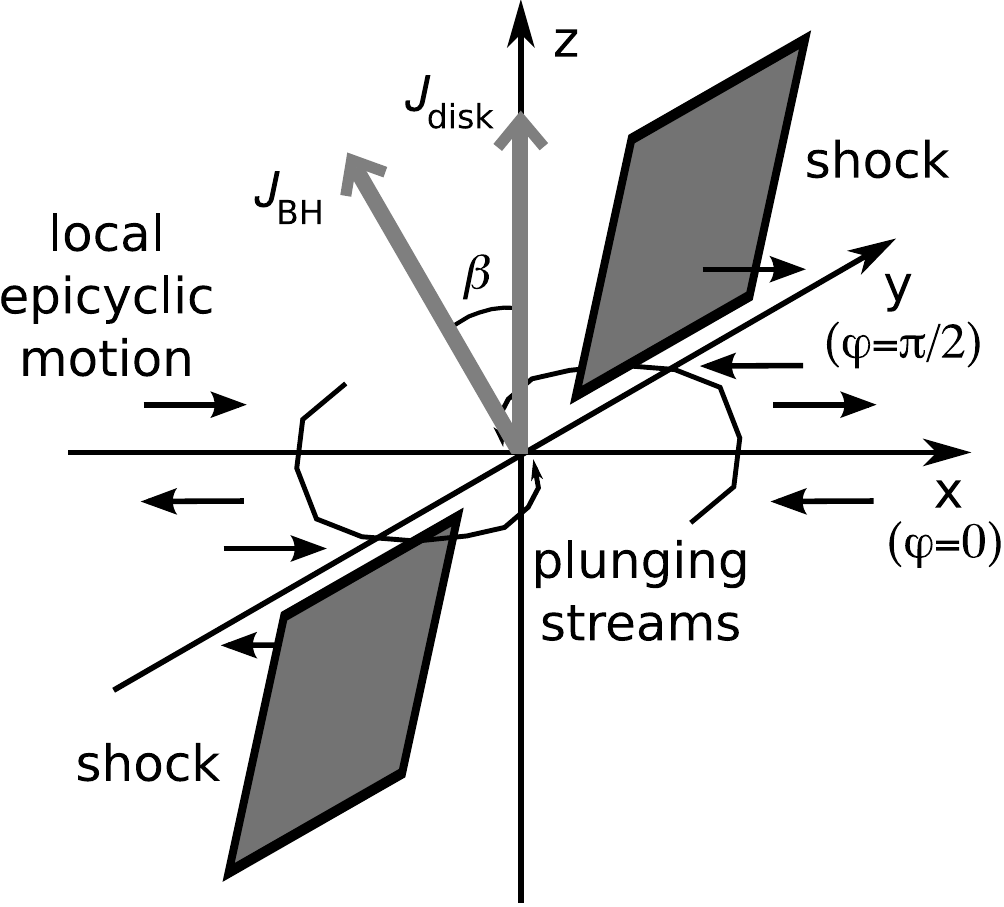} \caption{Schematic diagram of the inner
region of the tilted accretion disk, showing the pattern of
epicyclic motion, the standing shock, and the plunging streams.
\label{fig:schematic}}
\end{figure}

We finish with a figure (\ref{fig:schematic}) which provides a
visual summary of our findings. We note the overall consistency of
our interpretation of the results: 1) The observed epicyclic motion
is in agreement with expectations for warped disks in the wave-like propagation limit; and 2) The shocks are located near
the apocenters of the epicyclic motion, as one expects when the eccentricity increases with decreasing radius.

\begin{acknowledgements}
We would like to recognize Christopher Lindner and Joseph Niehaus
for their contributions to this work. We also thank Steve Balbus
and Gordon Ogilvie for very useful discussions and comments and the anonymous referee for suggested improvements.
PCF gratefully acknowledges
the support of a Faculty R\&D grant from the College of Charleston
and a REAP grant from the South Carolina Space Grant Consortium.
This work was supported under the following National Science
Foundation grants and programs: AST~0707624,
Partnerships for Advanced Computational
Infrastructure, Distributed Terascale Facility (DTF), and Terascale
Extensions: Enhancements to the Extensible Terascale Facility.
\end{acknowledgements}

\clearpage

\begin{thebibliography}{}

\bibitem[\protect\citeauthoryear{{Anninos}, {Fragile}, \&
  {Salmonson}}{{Anninos} et~al.}{2005}]{ann05}
{Anninos}, P., {Fragile}, P.~C.,  \& {Salmonson}, J.~D. 2005, \apj, 635, 723

\bibitem[\protect\citeauthoryear{{Bardeen} \& {Petterson}}{{Bardeen} \&
  {Petterson}}{1975}]{bar75}
{Bardeen}, J.~M.,  \& {Petterson}, J.~A. 1975, \apjl, 195, L65

\bibitem[\protect\citeauthoryear{{Caproni} et~al.}{{Caproni}
  et~al.}{2007}]{cap07}
{Caproni}, A., {Abraham}, Z., {Livio}, M.,  \& {Mosquera Cuesta}, H.~J. 2007,
  \mnras, 379, 135

\bibitem[\protect\citeauthoryear{{Caproni}, {Abraham}, \& {Mosquera
  Cuesta}}{{Caproni} et~al.}{2006}]{cap06}
{Caproni}, A., {Abraham}, Z.,  \& {Mosquera Cuesta}, H.~J. 2006, \apj, 638, 120

\bibitem[\protect\citeauthoryear{{Chakrabarti}}{{Chakrabarti}}{1985}]{cha85}
{Chakrabarti}, S.~K. 1985, \apj, 288, 1

\bibitem[\protect\citeauthoryear{{De Villiers} \& {Hawley}}{{De Villiers} \&
  {Hawley}}{2003}]{dev03b}
{De Villiers}, J.,  \& {Hawley}, J.~F. 2003, \apj, 592, 1060

\bibitem[\protect\citeauthoryear{{De Villiers}, {Hawley}, \& {Krolik}}{{De
  Villiers} et~al.}{2003}]{dev03c}
{De Villiers}, J., {Hawley}, J.~F.,  \& {Krolik}, J.~H. 2003, \apj, 599, 1238

\bibitem[\protect\citeauthoryear{{Fragile} \& {Anninos}}{{Fragile} \&
  {Anninos}}{2005}]{fra05b}
{Fragile}, P.~C.,  \& {Anninos}, P. 2005, \apj, 623, 347

\bibitem[\protect\citeauthoryear{{Fragile} et~al.}{{Fragile}
  et~al.}{2007a}]{fra07a}
{Fragile}, P.~C., {Anninos}, P., {Blaes}, O.~M.,  \& {Salmonson}, J.~D. 2007a,
  in proceedings of the 11th Marcel Grossmann Meeting on General Relativity
  (astro-ph/0701272)

\bibitem[\protect\citeauthoryear{{Fragile} et~al.}{{Fragile}
  et~al.}{2005}]{fra05a}
{Fragile}, P.~C., {Anninos}, P., {Gustafson}, K.,  \& {Murray}, S.~D. 2005,
  \apj, 619, 327

\bibitem[\protect\citeauthoryear{{Fragile} et~al.}{{Fragile}
  et~al.}{2007b}]{fra07b}
{Fragile}, P.~C., {Blaes}, O.~M., {Anninos}, P.,  \& {Salmonson}, J.~D. 2007b,
  \apj, 668, 417

\bibitem[\protect\citeauthoryear{{Fragile}, {Mathews}, \& {Wilson}}{{Fragile}
  et~al.}{2001}]{fra01a}
{Fragile}, P.~C., {Mathews}, G.~J.,  \& {Wilson}, J.~R. 2001, \apj, 553, 955

\bibitem[\protect\citeauthoryear{{Fragile}, {Miller}, \&
  {Vandernoot}}{{Fragile} et~al.}{2005}]{fra05c}
{Fragile}, P.~C., {Miller}, W.~A.,  \& {Vandernoot}, E. 2005, \apj, 635, 157

\bibitem[\protect\citeauthoryear{{Hawley}}{{Hawley}}{2000}]{haw00}
{Hawley}, J.~F. 2000, \apj, 528, 462

\bibitem[\protect\citeauthoryear{Ivanov \& Illarionov}{Ivanov \&
  Illarionov}{1997}]{iva97}
Ivanov, P.~B.,  \& Illarionov, A.~F. 1997, \mnras, 285, 394

\bibitem[\protect\citeauthoryear{{Kondratko}, {Greenhill}, \&
  {Moran}}{{Kondratko} et~al.}{2005}]{kon05}
{Kondratko}, P.~T., {Greenhill}, L.~J.,  \& {Moran}, J.~M. 2005, \apj, 618, 618

\bibitem[\protect\citeauthoryear{Krolik \& Hawley}{Krolik \&
  Hawley}{2002}]{kro02}
Krolik, J.~H.,  \& Hawley, J.~F. 2002, \apj, 573, 754

\bibitem[\protect\citeauthoryear{{Kumar} \& {Pringle}}{{Kumar} \&
  {Pringle}}{1985}]{kum85}
{Kumar}, S.,  \& {Pringle}, J.~E. 1985, \mnras, 213, 435

\bibitem[\protect\citeauthoryear{{Maccarone}}{{Maccarone}}{2002}]{mac02}
{Maccarone}, T.~J. 2002, \mnras, 336, 1371

\bibitem[\protect\citeauthoryear{{Miller} \& {Homan}}{{Miller} \&
  {Homan}}{2005}]{mil05}
{Miller}, J.~M.,  \& {Homan}, J. 2005, \apjl, 618, L107

\bibitem[\protect\citeauthoryear{{Miller} et~al.}{{Miller}
  et~al.}{2004}]{mil04}
{Miller}, J.~M., et~al. 2004, \apj, 601, 450

\bibitem[\protect\citeauthoryear{{Natarajan} \& {Armitage}}{{Natarajan} \&
  {Armitage}}{1999}]{nat99}
{Natarajan}, P.,  \& {Armitage}, P.~J. 1999, \mnras, 309, 961

\bibitem[\protect\citeauthoryear{Nelson \& Papaloizou}{Nelson \&
  Papaloizou}{1999}]{nel99}
Nelson, R.~P.,  \& Papaloizou, J. C.~B. 1999, \mnras, 309, 929

\bibitem[\protect\citeauthoryear{{Papaloizou} \& {Lin}}{{Papaloizou} \&
  {Lin}}{1995}]{pap95a}
{Papaloizou}, J.~C.~B.,  \& {Lin}, D.~N.~C. 1995, \apj, 438, 841

\bibitem[\protect\citeauthoryear{{Quataert}}{{Quataert}}{2003}]{qua03}
{Quataert}, E. 2003, Astronomische Nachrichten Supplement, 324, 435

\bibitem[\protect\citeauthoryear{{Shakura} \& {Sunyaev}}{{Shakura} \&
  {Sunyaev}}{1973}]{sha73}
{Shakura}, N.~I.,  \& {Sunyaev}, R.~A. 1973, \aap, 24, 337

\bibitem[\protect\citeauthoryear{{Torkelsson} et~al.}{{Torkelsson}
  et~al.}{2000}]{tor00}
{Torkelsson}, U., {Ogilvie}, G.~I., {Brandenburg}, A., {Pringle}, J.~E.,
  {Nordlund}, {\AA}.,  \& {Stein}, R.~F. 2000, \mnras, 318, 47

\end{thebibliography}

\end{document}